# Muon Cooling Progress and Prospects for an S-channel Muon Collider Higgs Factory

Mary Anne Cummings[1]

*Muons, Inc.*
*522 N Batavia Rd, Batavia, Il , 60510, USA*

Muon-based accelerators have the potential to enable facilities at both the Intensity and the Energy Frontiers. Muon storage rings can serve as high precision neutrino sources, and a muon collider is an ideal technology for a TeV or multi-TeV collider. Progress in muon accelerator designs has advanced steadily in recent years. In regard to 6D muon cooling, detailed and realistic designs now exist that provide more than 5 order-of-magnitude emittance reduction. Furthermore, detector performance studies indicate that with suitable pixelation and timing resolution, backgrounds in the collider detectors can be significantly reduced thus enabling high quality physics results. Thanks to these and other advances in design and simulation of muon systems, technology development, and systems demonstrations, muon storage-ring-based neutrino sources and a muon collider appear more feasible than ever before. A muon collider is now arguably among the most compelling approaches to a multi-TeV lepton collider and an S-Channel Higgs Channel.

---

[1] Work in part supported in part by the Department of Energy, under contract SBIR DE-SC0007634

Insert PSN Here

# 1. INTRODUCTION

I'll start with the charge from the P5 report [1]:

"For $e^+e^-$ Colliders the primary goals are:–Improving the gradient and lowering the power consumption". I will suggest here a thought experiment. Looking over the general parameters for $e^+e^-$ machines, we can see that both goals are achieved by increasing the mass of the electrons. The immediate consequence is that the higher mass electrons will have greatly suppressed radiation ($\sim m^{-4}$); enabling multi-pass acceleration; the gradient is improved by number of turns. Non-radiating electrons, with multi-pass acceleration, consume less power. So, that would be a substantial achievement. But how do we increase the electron mass? Well, generic electron, or lepton, masses are quantized in this universe into only three states of which we are aware: 0.511, 105.6 and 1777 MeV. The 1777 MeV is very hard to produce, and we wouldn't achieve the kind of bright beams needed for future colliders. The 105.6 MeV electron can be produced in sufficient quantities, thought its defuse production and finite lifetime presents a challenge – whose solutions will be presented below. The 105.6 MeV electron lifetime plus the suppressed radiation is optimal for the next generation $e^+e^-$ collider: the 2.2 μs lifetime: it only requires $E' \gg m_e/c\tau_e = 0.16$MV/m acceleration gradient, well within current technologies.

.

# 2. PHYSICS CASE

## 2.1. Features of the "heavy electron" beam

There are several advantages to the higher mass electron (or muon, if we are dare say). Larger couplings to Higgs-like particles - possible low energy Higgs Factory, or Z', through and s-channel production, highly suppressed in the the International Linear (lepton) Collider (ILC) or the Compact Linear Collider (CLIC). The increased mass tamps down on "beamstrahlung" radiation which results in a much narrower energy spread – particularly at TeV energies or higher, which enables higher precision measurements. The higher mass enables a large suppression of synchrotron radiation, which enables easier and more cost effective acceleration in rings and allows for much smaller machine footprints, which reduces costs and enables construction of frontier machines at existing labs – like Fermilab.

## 2.2. Physics at a "heavy electron" collider

The unique features of a heavy electon (muon) collider allows superior performance for a large range of energies. Below $\sqrt{s} < 500$ GeV, the superb energy resolution enables precision scans of Standard Model (SM) thresholds: $Z^0h$, W+W- and top pairs. The enhanced s-channel production makes the heavy $e^+e^-$ ($\mu^+\mu^-$) collider a superior Higgs factor. For instance, a 3 TeV muon collider (MC) (as we now say) can achieve resolutions of 95% of luminosity within an energy range dE/dx ~ 0.1%. This compares with only 35% luminosity contained within an energy range dE/dX of ~ 1%. Figures 1a and 1b illustrate the enhanced physics capability of high resolution heavy lepton machine. There are theoretical speculations that the LHC Higgs signal may be a combination of two nearly degenerate Higgs-like bosons. This could happen in some models, for example, two Higgs doublet models. plus 1 singlet models,

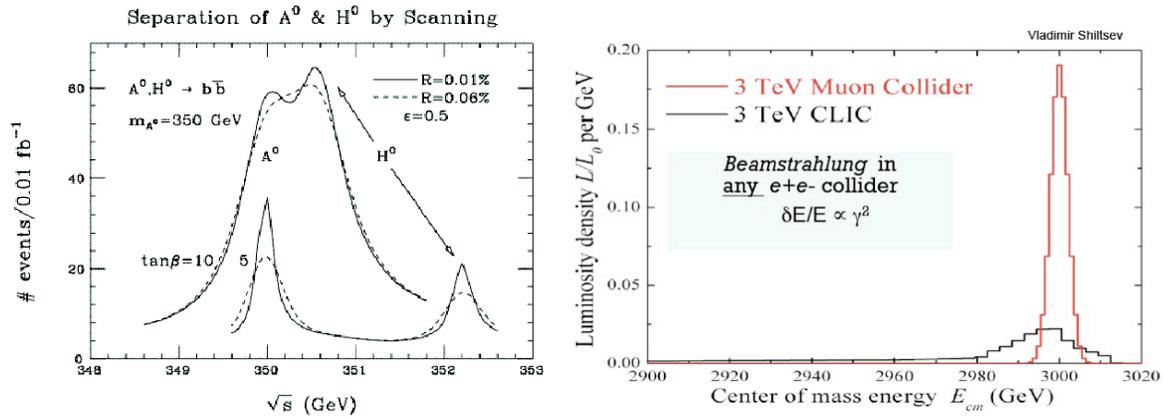

Figure 1a: (Left) Separation of minimum super-symmetric closely degenerate $A^0, H^0$ Higgs states possible with MC resolutions [2]. Figure 1b: (Right) Comparison of 3 TeV MC and CLIC resolution performance [3].

A putative1 GeV level degeneracy could be easily resolved at an early stage of the muon collider program, when the Higgs mass window is determined. However, we can go much further, and 1 MeV level mass degeneracy resolution achievable at a heavy electon collider. As an example, Figure 2a shows a mass plot where the interference effect between two highly degenerate Higgs bosons has to be taken into account.

For $\sqrt{s} > 500$ GeV, this resolution makes the MC sensitive to Beyond Standard Model Searches. high luminosity is required for these searches. Cross sections for central ($|\theta| > 10^o$) region pair production, are ~ 86.8 fb/s(in TeV$^2$) where (R ≈ 1). For example, at $\sqrt{s}$ =3.0 TeV, we would expect 965 events/unit of R, where R ≡ $\sigma/\sigma_{QED}$ ($\mu+\mu-$-->$e+e-$) ~ 1. Figure 2b shows typical pair production cross sections of interest.

Because of their compact nature and easy acceleration, multi-TeV capability for energy frontier machines is a long-term possibility for heavy electron colliders. Designs have matured for luminosities > $10^{34}$ cm$^{-2}$ s$^{-1}$. There is also an option for 2 interactions points. For $\sqrt{s} > 1$ TeV, fusion processes become important, as illustrated in Figure 2c. The muon collider becomes and effective Electroweak Boson collider. At $\sqrt{s}$ ~ 10 TeV it has a similar discovery reach as a 100 TeV pp collider, and would be a complementary discovery machine. At $\sqrt{s} > 5$ TeV, Higgs self-coupling can be measured with a resolution < 10%.

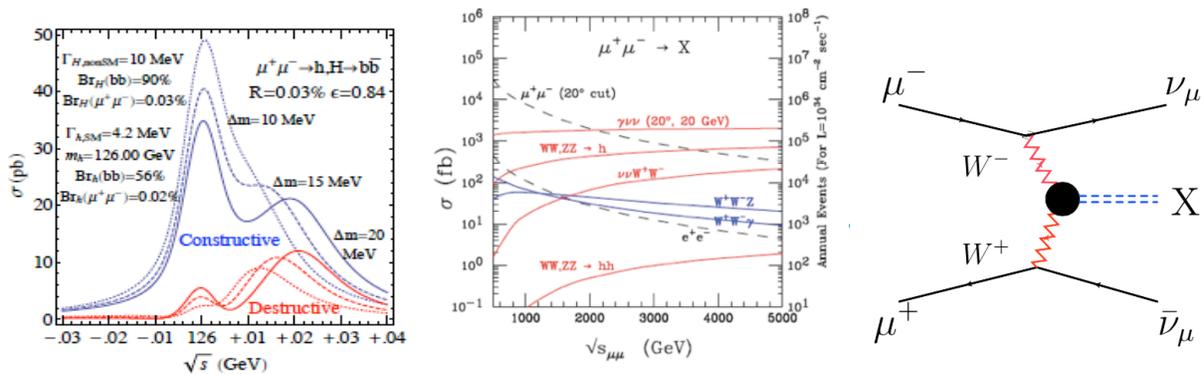

Figure 2a: (Left) Resolving highly degenerate Higgs bosons at the muon collider through scanning[4]. Figure 2b: (Center) Pair production cross sections for a variety of SM interactions to search for beyond SM physics [3]. Figure 2c: (Right) Vector Boson Fusion reactions, common for multi-TeV MC.

Insert PSN Here

## 3. CHALLENGES

While the advantages for a heavy electron collider are compelling there are challenges, though none are now thought to be "show-stoppers". Two features of these beams have to be addressed: 1) Production is into a diffuse phase space. Heavy electrons (muons) are tertiary beams from p → π → μ ν which need to be captured and focussed quickly. 2) Heavy electrons have short lifetimes, so their acceleration must be rapid. The backgrounds from decays in the colliding ring have to be dealt with, and at multi-TeV energies, neutrino radiation becomes a concern. For the latter, accelerator and detector technologies have evolved such that are sufficient to address these issues. For the former, ionization cooling is the only effective method to quickly capture and focus heavy electrons, and the canard has been that this requires a new and "unproven" technology. But this method is conceptually not new and the technology to effect cooling channels has been demonstated which I describe below.

### 3.1. The Issue of Cooling

All high-energy lepton colliders need transverse cooling. For example, for "light electron" accelerators, radiation damping is used to focus these electrons: momentum is lost through a bend due to synchrotron radiation, and regained only in the forward direction with radiofrequency (RF) acceleration. For heavy electrons, with finite lifetimes, ionization cooling is optimal. Instead of a magnetic bend, these heavy electrons traverse through a low-Z absorber, losing momentum in all three dimensions, and as in the radiation damping methods mentioned above, momentum is restored the forward or "z" direction with RF. This is shown in Figures 3a and 3b.

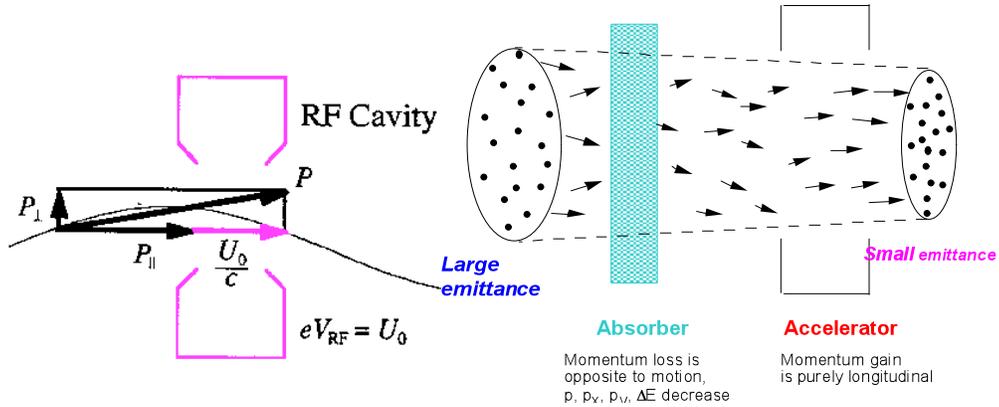

Figure 3a. (Left) Light electron cooling (radiation damping): momentum is lost in the magnetic bend via sychrotron radiation and restored with RF. Figure 3b. (Right) Heavy electron cooling (ionaization): momentum is lost in absorber material (dE/dx) and restored with RF.

### 3.2. Ionization Cooling and Emittance Evolution.

The cooling performance is measured in the change of the beam emittance, a measure for the average spread of particle coordinates in position-and-momentum phase space. Equation 1 describes the emittance evolution through a typical cooling channel.

$$\frac{d\varepsilon_N}{ds} = -\frac{1}{P_l}\frac{dP_l}{ds}\varepsilon_N + \frac{\beta_\perp E_S^2}{2\beta^2 m_l c^2 L_R E} \qquad (1)$$

Here, β is the relativistic velocity and $β_⊥$ is the beam size defined by the strength of the magnetic channel, $L_R$ is the radiation length of the absorber material. The evolution has a cooling term (ionization) and a heating term (multiple scattering). The choice of absorber material and magnetic channel can minimize the heating effect, so absorbers are low Z materials (typically hydrogen). Ionization cooling is inherently transverse, and this often comes at the expense of longitudinal beam heating. To counter this, and to achieve the complete, "six dimensional" cooling (for the three dimensional canonical variables, position and momentum) emittance exchange is employed, which exploits a beam "dispersion" (relation of beam position to momentum), in order to exchange longitudinal emittance for transverse emittance, and with subsequent transverse cooling. Illustrations of these techniques are shown in Figures 4a and 4b.

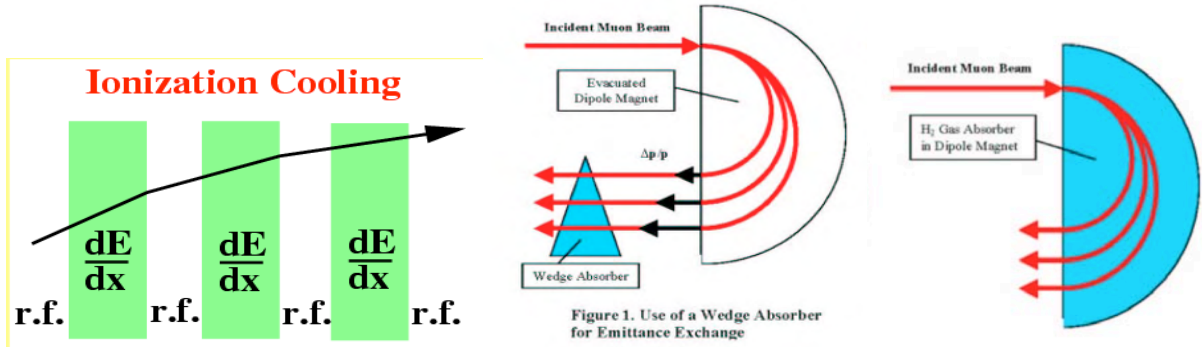

Figure 4a) (Left) Transverse cooling channel for "4-D" cooling, suitable for Neutrino Factories. Figure 4b) (Center, right) Emittance exchange to effect "6-D" cooling needed for Colliders, for a discrete and continuous absorbers.

Combinations of transverse cooling and emittance exchange, as well as bunch merging (for colliders) give rise to emittance evolution plots (one example is shown in Figure 5) for cooling channels that accommodate a Neutrino Factory (heavy electron storage ring sources of neutrinos), Higgs Factory (lower luminosity for enhanced s-channel production) and Energy Frontier Collider. For the high luminosity colliders, a "reverse emittance exchange" (final cooling) is employed where transverse emittance is exchanged into longitudinal emittance. Current collider ring designs can accommodate longitudinal emittances of 3 mm mr or larger, for larger momentum spreads acceptable for Energy Frontier machines. Note that the front ends for Neutrino Factory, Higgs Factory and multi-TeV colliders are the same.

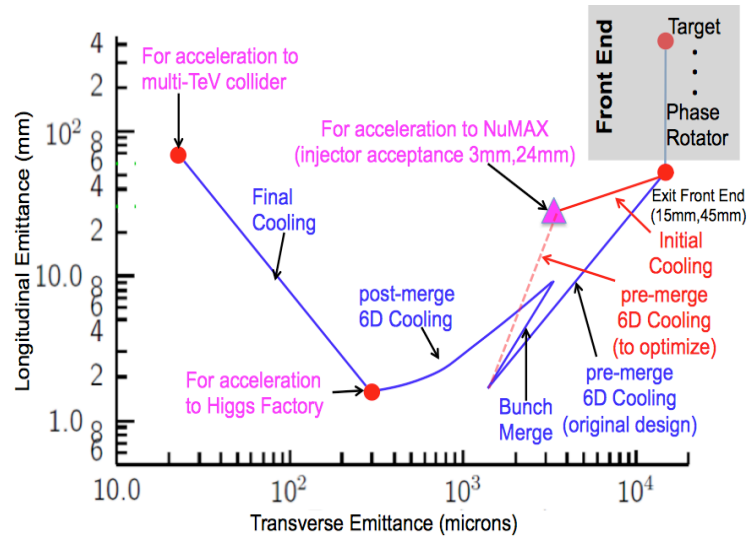

Figure 5: Emittance evolution in a cooling channel.

Insert PSN Here

## 4. MUONS, INC. INNOVATIONS

Three novel ideas have been developed at Muons, Inc. that have enhanced the viability of heavy electron colliders:

1. The idea of a gaseous energy absorber enables new technology to generate high accelerating gradients for muons by using the high pressure region of the Paschen curve: High Pressure RF Cavity (HPRF) TESTED!
2. Concept of a cooling channel filled with continuous homogenous absorber to provide longitudinal cooling by exploiting the path length correlation with momentum in a magnetic channel with positive dispersion: Helical Cooling Channel (HCC). COMPLETE END-END SIMULATION DONE
3. Parametric Resonance Ionization Cooling (PIC) concepts are being developed that can decrease emittance by factor of 10 beyond the "equilibrium emittance" of current cooling techniques. STILL DEVELOPING…

### 4.1. Pressurized RF Cavities

A gaseous energy absorber enables an entirely new technology to generate high accelerating gradients for muons by using the high-pressure region of the Paschen curve. This idea of filling RF cavities with gas is new for particle accelerators and is only possible for heavy electrons (muons) because they do not scatter as do strongly interacting protons or shower as do less-massive electrons. Measurements by Muons, Inc. and IIT at Fermilab have demonstrated that hydrogen gas suppresses RF breakdown very well, about a factor six better than helium at the same temperature and pressure [5]. Consequently, much more gradient is possible in a hydrogen-filled RF cavity than is needed to overcome the ionization energy loss, provided one can supply the required RF power. Hydrogen is also twice as good as helium in ionization cooling effectiveness, viscosity, and heat capacity. Tests were preformed of materials in pressurized RF cavities in magnetic fields where an external field causes no apparent reduction in maximum achievable gradient. The RF cavity tested and the results are shown in Figures 5a and 5b.

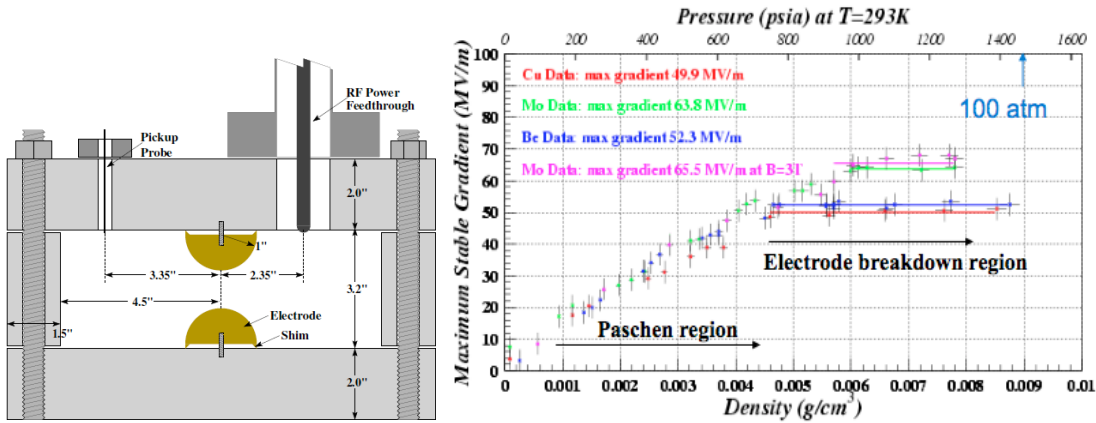

Figure 5a: (Left) Schematic for a high-pressure RF cavity. Figure 5b: (Right) Results from tests that confirm the Paschen curves for various metals. These were done inside a 3.5 Tesla magnetic field with no affect on the gradients.

The critical test of this concept for breakdown suppression and for continuous absorber cooling channel technology was a beam test performed at the Muon Test Area at Fermilab with the full power of the Fermilab 400 MeV Linac beam. The major concern was the loading of the cavity by beam-induced plasma and whether the beam would induce electric breakdown. The answer is NO [6]! But the beam-loading effect on the cavity can be moderated by using a 0.2% mixture of O2. Figures 6a and 6b illustrate the results.

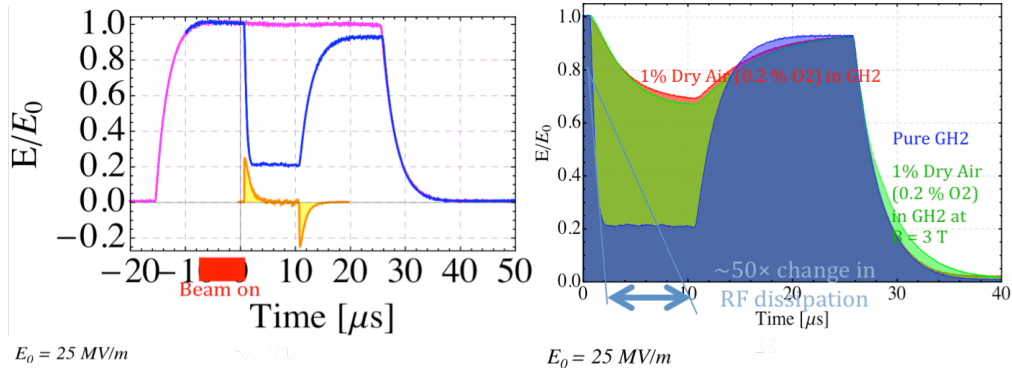

Figure 6a: Observed RF pickup signal and toroid monitor signal (measuring beam current) in the GH2 filled RF cavity. Significant RF amplitude drop is seen when the beam is on. Figure 6b: Constrast in beam loading effects between in RF cavity filled with GH2 only, and with 0.2% O2 ("dry air").

### 4.2. Helical Cooling Channel (HCC)

The HCC is an attractive example of a cooling channel based on the idea of energy loss dependence on path length in a continuous absorber. One version of the HCC uses a series of high-gradient RF cavities filled with dense hydrogen gas, where the cavities are in a magnetic channel composed of a solenoidal field with superimposed helical transverse dipole and quadrupole fields [7]. In this scheme, energy loss, RF energy regeneration, emittance exchange, and longitudinal and transverse cooling happen simultaneously. The helical dipole magnet creates an outward radial force due to the longitudinal momentum of the particle while the solenoidal magnet creates an inward radial force due to the transverse momentum of the particle, or

$$\begin{aligned} F_{h\text{-}dipole} &\cong p_z \times B_\perp; \quad b \equiv B_\perp \\ F_{solenoid} &\cong p_\perp \times B_z; \quad B \equiv B_z, \end{aligned} \quad (2)$$

where $B$ is the field of the solenoid, the axis of which defines the $z$ axis, and $b$ is the field of the transverse helical dipole at the particle position. By moving to the rotating frame of the helical fields, a time and z- independent Hamiltonian can be formed to derive the beam stability and cooling behavior [7]. Use of a continuous homogeneous absorber as shown on the right side of Figure 4c, rather than wedges at discrete points, implies a positive dispersion along the entire cooling path, a condition that has been shown to exist for an appropriately designed helical dipole channel. We have also shown that this condition is compatible with stable periodic orbits. The simple idea that emittance exchange can occur in a practical homogeneous absorber without shaped edges followed from the observation that RF cavities pressurized with a low Z gas are possible, described above. The analytic relationships derived from this analysis were used to guide simulations using a code developed based on the GEANT4 based program called G4Beamline [8] and also using ICOOL developed at BNL. Current simulations with four segments of HCCs (adjusting field and sizes for an increasingly cooled and focused heavy electron beam) with realistic matching show ~ 5 X $10^5$ reduction in the six dimensional emittance. close to the cooling necessary for an energy frontier collider with a luminosity of ~ $10^{34}$ cm$^{-2}$ s$^{-2}$. The orbits from these simulations are illustrated in Figure 6a.

Insert PSN Here

## 4.3. HCC Engineering design

The current engineering design of a helical cooling channel employs individual helical solenoid (HS) coils displaced in a helical pattern to create the necessary dipole and quadrupole components. RF cavities are imbedded inside the solenoids. Figures 6b and 6c illustrate the arrangement of these combined magnets and RF cavities[9]. The key technologi-cal barriers include cooling of the helical solenoid coils made of the low temperature superconductor (LTS) in the presence of RF cavities embedded in the channel and operation of RF cavities in the presence of a magnetic field. Recent results described above show pressurizing the RF cavities is one solution to the latter challenge. The first challenge, we believe, can be addressed by reducing the size of each RF cavity utilizing a low loss dielectric insert to ease physical constraints for a given frequency. This allows enough room between the cavities and the coils for the magnet coils, the magnet cryostat, the hydrogen pressure vessel, and the RF coaxial feeds to the individual cavities. Calculations show that the heat loads will be tolerable and RF breakdown of the inserts will be suppressed by the pressurized hydrogen gas. The work done on this problem is one of the key accomplishments of the project and has led to a solution to the engineering problem of feeding the RF power through the magnet cryostat. The current design build is the 10 T, 805 MHz segment of a HCC. This corresponds to the second section of four HCC segments of decreasing size and increasing magnetic fields to accommodate an increasingly cooled heavy electron beam in a complete cooling channel.

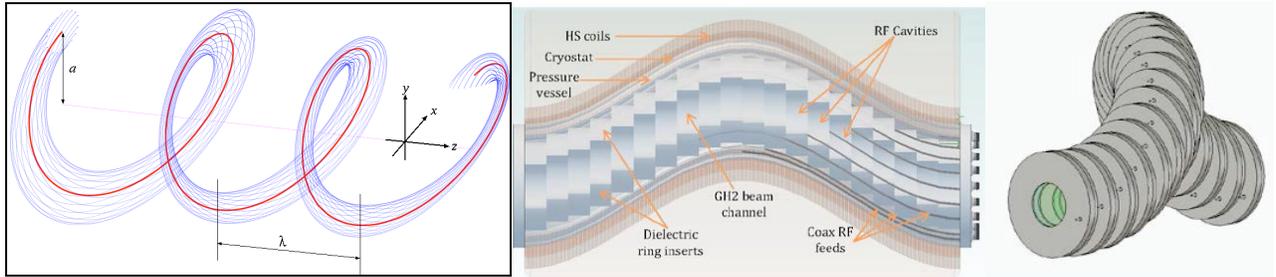

Figure 7a: (Left) Simulation of orbits in an HCC. Figure 7b: (Center) Conceptual diagram of an HCC module showing dielectric-loaded RF cavities enclosed in a pressure vessel that "screws" in to the HS cryostat. Figure 7c: (Right) Another view of RF cavities aligned to fit inside an HS magnet and cryostat.

## 4.4. Extreme Cooling: PIC and HCC

To go beyond the limits of ionization cooling described above, we need to consider more sophisticated manipulations of phase space. One idea is to combine ionization cooling with parametric resonances that is expected to lead to heavy electron beams with much smaller transverse sizes [10]. We studied a linear magnetic transport channel where a half integer resonance is induced such that the normal elliptical motion of particles in $x$-$x'$ phase space becomes hyperbolic, with particles moving to smaller $x$ and larger $x'$ at the channel focal points. Thin absorbers placed at the focal points of the channel then cool the angular divergence of the beam by the usual ionization cooling mechanism where each absorber is followed by RF cavities. To deal with the complicated aberrations, combined HCC fields with one at opposite helicity to create alternating dispersions – and cancellations of some of the aberrations. Recently, a model using skew quadrupoles to mix both X and Y planes. Figures 7a and 7b illustrate the Parametric Resonant Ionization (PIC) concept and give a possible lattice configuration for its implementation. This is a very challenging and intriguing problem, which shows great promise in the potential to gain another factor of ten cooling factor over current ionization cooling channels.

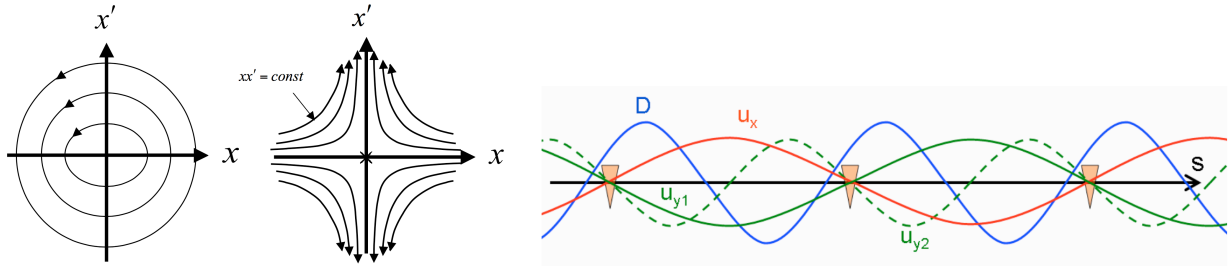

Figure 8a: (Left) PS area is reduced in x due to the dynamics of the parametric resonance and reduced in x' by ionization cooling. Figure 8b: (Right) Lattice functions of a beam cooling channel suitable for PIC showing the dispersion D, the horizontal betatron amplitude $u_X$ and two possible solutions for the vertical betatron amplitude, $u_{Y1}$ and $u_{Y2}$. The triangles represent the wedge absorbers. The dipoles and quadrupoles are not shown.

## 5. HEAVY LEPTON ACCELERATION PROGRAM AT FERMILAB

Heavy electrons do require an ultrafast accelerator chain, whose operating parameters are just beyond current "standard machines", but will be accommodated by new accelerators currently under study. Recirculating Linear Accelerators (RLAs), Fixed-Field Alternating-Gradient (FFAG) Rings and Rapid Cycling Synchrotrons (RCS) and Superconducting Linacs are all candidates for various configurations of heavy electron machines, the latter two being studied and developed at Fermilab.

The unique synergies between current and near future programs at Fermilab, and the requirements of an array of possible future bright heavy electron machines makes Fermilab and ideal place to consider staging this ambitious program. For the sake of brevity, Figure 9 gives a diagram of possible machine staging, from storage ring sources of neutrinos to Higgs Factory and Energy Frontier Collider. The compact nature of these machines not only keeps costs down, but can easily fit on the current Fermilab site.

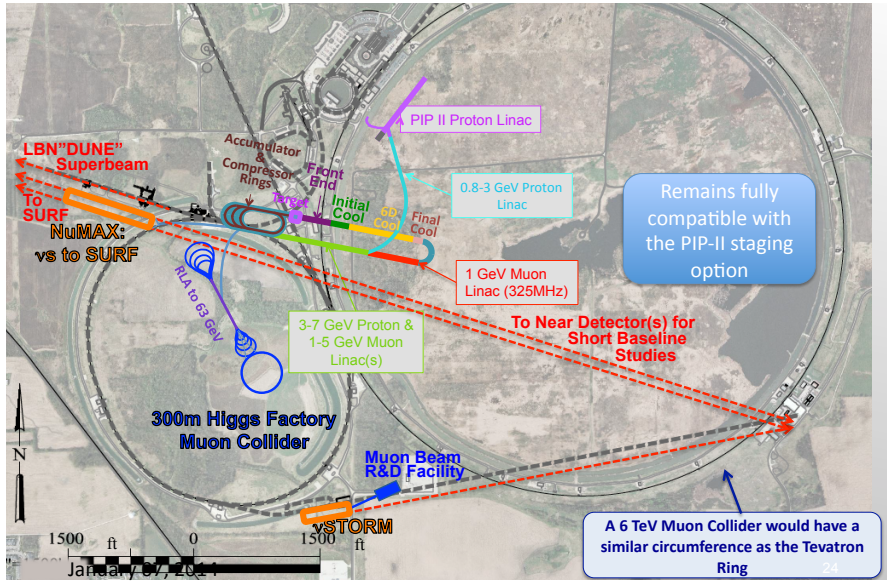

Figure 9: Layout of staging scheme for series of possible heavy electron (muon) machines.

Insert PSN Here

## 6.  PARTING THOUGHTS – THINKING MORE BROADLY

Despite (or, consistent with) the current P5 priorities, a staged muon (or, heavy electron) acceleration program at Fermilab is a viable path to a Higgs factory and a possible energy frontier future, starting with what is already planned and present in the current program. The canard that ionization cooling is an "unproven technology" has lost traction. Cooling is a standard practice in accelerator science, ionization is settled physics and the **essential technology** for ionization cooling has now been proven **viable**. Necessary accelerator and magnetic field gradients have already been demonstrated. Some elements of these future machines could be implemented in the current "future" LBNE-DUNE neutrino program. at Fermilab. A muon storage ring source of neutrinos could be a possible upgrade. This would be consistent with the international character of our aspirations at Fermilab, and a broad long-term future for particle physics in the U.S. in general. In particular, Muons, Inc. innovations in cooling and other accelerator technologies are making bright muon (heavy electron) beams (and other types of electron beam) more **affordable** and **feasible**. These possibilities do not detract from the current, necessarily focused program at Fermilab. There are great synergies for upgrades to currently planned neutrino and muon programs. Staged implementation of these new accelerators would give significant cost savings, and make future energy frontier machines possible.

## Ackowledgments

Work supported by Department of Energy SBIR contract DE-SC0007634